\documentstyle[11pt,psfig]{article}
\newcommand{\EP}{$\rm p(e,e'K)\Lambda$\ }
\newcommand{\PP}{$\gamma\rm p \longrightarrow \Lambda K^+$\ }
\begin{document}

\title{Models for photo- and electro-production
of K$^+$ \\ in view of new data}

\author{P. Byd\v{z}ovsk\'y$^1$\footnote{E-mail: bydz@ujf.cas.cz}, 
F. Cusanno$^2$, S. Frullani$^2$, F. Garibaldi$^2$, M. Iodice$^3$, 
M. Sotona$^1$,\\ and G.M. Urciuoli$^2$\\
 \\
{\it $^1$Nuclear Physics Institute, \v{R}e\v{z} near Prague, 
the Czech Republic} \\
{\it $^2$INFN, Gruppo Sanit\`{a} and Instituto Superiore di
Sanit\`{a}, Physics Lab, Rome, Italy} \\
{\it $^3$INFN Sezione di Roma III, Rome, Italy}}
\maketitle
\begin{abstract}
Predictions of isobaric and Regge models are compared with 
the latest experimental data (Bonn, JLab) to select among 
the models those providing a satisfactory description of 
the data. Only the Saclay-Lyon (SLA) and MAID models are 
in a reasonable agreement with both photo- and 
electro-production data ranging up to about 
$E^{lab}_{\gamma}=2.5$ GeV. In this energy region the Regge 
model is suitable only for description of the 
electro-production data and the older models Adelseck-Saghai 
(AS1, AS2), Workman (W1,W2), and Adelseck-Bennhold (AB2) can 
reliably predict only the photo-production cross sections up 
to $E^{lab}_{\gamma}=1.5$ GeV. In the kinematic region of 
the E98-108 experiment on the electro-production, SLA, 
Adelseck-Wright (AW4), Regge and MAID models are expected to 
provide an appropriate description of the separated cross 
sections whereas the unseparated cross sections can be 
reasonably well predicted also by the AW3 and 
Williams-Ji-Cotanch (C4) models.
\end{abstract}

\section{Introduction}

The photo- and electro-production of kaons on the proton 
in the resonance region have been studied both experimentally 
and theoretically since the 1960s. Many data were collected 
on the kaon photo-production (Cornell, Cal Tech, Bonn, Tokyo, 
DESY, and Orsay, see \cite{AS90}) but only a few experiments 
were carried-out on the electro-production (Harvard-Cornell, 
DESY, and Cambridge, see \cite{Beb77} and references therein). 
Simulated by the availability of new facilities with higher 
current or/and high duty factor and polarization capability 
(JLab, Bonn, Grenoble), more precise data on both reactions 
have been accumulated starting in the late of 1990s which 
renewed interest in the subject. Now various response 
functions are accessible and measured with a good level of 
precision. The latest experimental data, especially those 
from JLab \cite{Nic98,Moh02,E98} on the separated cross 
sections in the electro-production in the kinematic region 
where scarce data were formerly available, allow to perform 
more rigorous tests of theoretical models and in this way 
improve our understanding of the elementary process. 

Numerous theoretical attempts have been made to describe 
the electro-magnetic production of kaons with $\Lambda$ 
in the final state. In the kinematic region, 
$E^{lab}_{\gamma}=0.91-2.5$ GeV, the isobaric 
models \cite{AS90,ABW85,AW88,W91,WJC91,WJC92,SL96,SLA98} 
are of particular interest.
In these models the amplitude is derived from an effective
hadronic Lagrangian using the Feynman diagram technique in 
the tree level approximation.
However, it has been shown that the new data on \EP 
\cite{Nic98} can be equally well described by the Regge
model \cite{Rgg97} which is based on the idea of exchanges
of families of the particles with particular quantum numbers 
in the t-channel. The Regge model was aimed mainly for
higher energies ($E_{\gamma}^{lab} > 4$ GeV) and small angles. 
However, the model was successfully applied to description of 
the electro-production data at the centre of mass energy 
$W=1.84$ GeV \cite{Rgg97}.

In the isobaric models the invariant amplitude gains 
contributions from the extended Born diagrams, in which 
the proton, $\Lambda$, $\Sigma^0$, and kaon are exchanged 
in the intermediate state, and the resonant diagrams which 
include exchanges of moderate mass (less than 2 GeV) 
nucleon, hyperon, and kaon resonances.
Unfortunately, due to absence of a dominant exchanged baryon 
resonance in the process \cite{Tho66}, in contrast with the 
pion and eta production, many of exchanged resonances have 
to be {\it a priori} assumed to contribute \cite{AS90,SL96} 
introducing a rather large number of free parameters in 
calculations, the appropriate coupling constants.
The free parameters are determined by fitting the cross 
sections and polarizations to the experimental data which, 
however, provides a copious number of possible sets of 
parameters \cite{AS90,SL96}. This large number of models 
which describe the data equally well can be reduced 
implementing duality hypothesis, crossing symmetry, and 
SU(3) symmetry constrains.

According to duality principle most of the nucleon resonances
exchanged in the s-channel, especially those with a high spin,
can be mimic by the lowest-mass kaon poles K$^*$ and K$_1$
in the t-channel \cite{WJC92,SL96}.
The crossing symmetry constraint requires that a realistic 
model must yield simultaneously a reasonable description of 
the radiative capture of K$^-$ on the proton with the lambda 
in the final state, which is related to the \PP via the 
crossing symmetry \cite{WJC92,SL96}. 
The flavor SU(3) symmetry allows to relate the main coupling
constants $g_{\rm K\Lambda N}$ and $g_{\rm K\Sigma N}$
to the better established one,  $g_{\pi\rm NN}$.
For the 20\% breaking of the SU(3) symmetry the following 
limits can be obtained:
$-4.4 \leq g_{\rm K\Lambda N}/\sqrt{4\pi} \leq -3.0$ and
$0.8 \leq g_{\rm K\Sigma N}/\sqrt{4\pi} \leq 1.3$  
\cite{AS90,SL96}. Analysis of data performed under 
different assumptions about duality, crossing and SU(3) 
symmetry \cite{AS90,ABW85,AW88,W91,WJC92,SL96,SLA98} showed 
that a moderate number of resonances is sufficient to get 
a reasonable agreement with the experimental data.

The models discussed above assume point-like particles  
in the hadron vertexes to ensure the gauge invariance 
principle. Recently Haberzettl {\it et al.} 
\cite{HBMF98,Ben99} introduced hadron form factors in 
the gauge-invariant way in an isobaric model which 
required addition of a contact term to compensate 
the gauge-violating part of the amplitude. The method 
was worked out recently by Davidson and Workman \cite{DW01} 
and further used by Janssen {\it et al.} \cite{Jan01}.  
Taking into account the hadron form factors led to 
reducing of divergences at higher energies inherent to 
most of isobaric models and also to more realistic 
predictions for the isospin-symmetry related 
$\rm n(\gamma,K^0)\Lambda$ channel \cite{HBMF98}.

Another simplification assumed by the models is neglecting   
a meson-baryon re-scattering in the final state which 
obviously leads to violation of unitarity.  
Interaction of hadrons in the final state was taking into 
account for the photo-production reaction by Feuster and 
Mosel \cite{FM98} employing the K-matrix approximation. 
Enforcing unitarity dynamically was performed by 
Chiang {\it et al.} \cite{Ch01} who utilised the 
coupled-channel approach. In their analysis they concluded 
that inclusion of the $\pi$N system in the intermediate 
state is needed for a proper description of the \PP reaction. 

More elementary approaches to study the reaction mechanism 
of \PP was performed in terms of quark degrees of freedom 
in Refs. \cite{ZpL95,LLP95,FHZ91}. These models being in 
a closer connection with QCD than those based on the hadron 
degrees of freedom, need a smaller number of parameters to
describe the data. Moreover, the quark models assume explicitly 
an extended structure of the hadrons which was found to be 
important for a reasonable description of the photo-production 
data \cite{LLP95}. Other approach to the \PP reaction based 
on the Chiral Perturbation Theory \cite{Chpt} is applicable 
to the threshold region only.

This paper is aimed to discuss existing isobaric and Regge 
models for the electro-magnetic production of K$^+$ off 
the proton particularly in view of the new experimental 
data \cite{Nic98,Moh02,E98,SPH98}. 
We intend to pick up among the models those reliable in the 
kinematic region of the E98-108 experiment \cite{E98} for 
further comparison with the coming data. The experiment 
E98-108 which was carried out in Hall A at Jefferson Lab 
in 2001 and 2002 extends the present data on the separated 
cross sections of the \EP reaction to higher energies and 
larger $Q^2$ ($Q^2=-q^2_{\gamma}$) providing very good 
quality data and consequently it can set stringent 
constraints on the models. 
In the next section we briefly outline the models. 
In the section 3 we discuss predictions of the models in 
comparison with the new data.

\section{Models}

In comparison performed here we adopted a set of isobaric
models, which are sufficiently well described in literature 
to enable performing calculations, and the Regge model. 
We also adopted results of the model by Bennhold {\it et al.}  
utilising the ``new isobar model (Dec 2000)'' interactive 
code at the KAON-MAID Web-page \cite{Ben99}. 
Those, trivial, isobaric models which assume only the Born 
terms (AB1 \cite{ABW85} and AW1 \cite{AW88}) were omitted 
from the analysis. 

The isobaric models selected for the comparison can be 
grouped according to the reaction they were determined for:\\
\ 1. photo-production only: AB2 \cite{ABW85}, AW2 \cite{AW88}, 
AS1, AS2 \cite{AS90}, W1 and W2 \cite{W91}.\\
\ 2. electro-production only: AW3 \cite{AW88}.\\
\ 3. photo- and electro-production: AW4 \cite{AW88},
C3 \cite{WJC91}, C4(WJC) \cite{WJC92}, SL \cite{SL96} and SLA
\cite{SLA98}.

The models in the first group have to be extended for \EP 
assuming a particular prescription for the electro-magnetic 
form factors of the mesons and baryons. Results of these 
models for the electro-production, therefore, depends to 
some extent on an ansatz we made for the electro-magnetic 
form factors and in this limited sense the models are 
predictive. In this analysis we used phenomenological 
prescriptions \cite{SF94} fixed by the electron scattering. 
The meson form factors were supposed to be of the monopole 
form with $m_V=0.77$ GeV for all mesons whereas the dipole 
form, with $G_E(Q^2)=(1+Q^2/m_V^2)^{-2}$ 
and $m_V=0.84$ GeV, was used for all baryons.

Unfortunately, there are several models, AW3, AW4, C3, 
and C4, which were not sufficiently well commented in 
the original paper and for which either a definition 
or a form of the electro-magnetic form factors is not 
clear. In that cases we tried several possibilities 
to introduce the form factors. Finally we have chosen 
those which provide the best agreement with
original results. That is why in the case of AW3 and AW4 
models we employ the phenomenological prescriptions as 
used in Ref. \cite{SF94} and discussed above. 
In the case of the Williams-Ji-Cotanch models, C3 and 
C4 (WJC), we find it interesting to compare the original
prescription in which the electric ($G_E$) and magnetic ($G_M$) 
form factors are used in the electro-magnetic vertexes, we 
denote it as C3\_4 and C4\_4 hereafter, with the definition  
commonly used \cite{SL96,SF94} in which Dirac ($F_1$) and 
Pauli ($F_2$) form factors are substituted in a 
gauge-invariant way \cite{SF94}, we denote it as C3\_2 and 
C4\_2. In both cases the Gari and Krumpelmann prescription, 
the extended vector meson dominance model \cite{GK92}, was 
utilised for the baryon form factors and the vector meson 
dominance model as discussed by 
Williams {\it et al.} \cite{WJC92} was used for mesons. 

In Tables~1 and 2 coupling constants are listed for the 
isobaric models adopted here. The effective parameters 
$G_x$, e.g. $G_{\rm N^*}=\kappa_{\rm NN^*}\,\cdot\, 
g_{\rm K\Lambda N^*}/\sqrt{4\pi}$ for the baryon exchanges 
and $G_{\rm K^*}^V=g_{\rm K^*K\gamma}\,\cdot\, 
g_{\rm K^*\Lambda p}^V/4\pi$ for the meson ones, 
are shown except for the main coupling constants for which 
$G_{\rm K\Lambda p}=g_{\rm K\Lambda p}/\sqrt{4\pi}$ 
and $G_{\rm K\Sigma p}=g_{\rm K\Sigma p}/\sqrt{4\pi}$ 
are stated. Values of the parameters can be directly 
compared since they are properly normalised according 
to definition of vertexes and units used here. Since the
observable quantities are bilinear forms of the CGLN 
amplitudes which are linear in the strong coupling constants 
in the tree level approximation the whole set of 
the constants may be multiplied by an arbitrary global phase, 
e.g. -1. In this way we can arrive with the same sign for 
the main coupling constant $G_{\rm K\Lambda p}$ which then 
makes a comparison more transparent.
%
% Table 1
%
\begin{table}[hbt]
\caption{Coupling constants of the isobaric models for the
photo-production of kaons. Some of the sets of the coupling
constants were multiplied by -1. to get the negative sign at
$G_{\rm K\Lambda p}$ .} \label{models1}
\smallskip
\begin{center}
\begin{tabular}{|lrrrrrr|}
\hline\vspace{1mm} Model    &AB2 \cite{ABW85} &AW2 \cite{AW88}
&W1 \cite{W91}& W2 \cite{W91}  &  AS1 \cite{AS90} & AS2 \cite{AS90} \\
\hline
$G_{\rm K\Lambda p}$ &-1.03 &-4.30 &-1.50 &-1.73 &-4.17 &-4.26 \\
$G_{\rm K\Sigma p}$  & 0.41 &-1.82 &-0.28 & 0.09 & 1.18 & 1.20 \\
$G^{V}_{\rm K^*}$    &-0.22 &-0.12 &-0.26 &-0.26 &-0.43 &-0.38 \\
$G^{T}_{\rm K^*}$    & 0.05 & 0.34 & 0.18 & 0.32 & 0.20 & 0.30 \\
$G^{V}_{\rm K1}$     &      &-0.27 &      &      &-0.10 &-0.06 \\
$G^{T}_{\rm K1}$     &      &-0.83 &      &      &-1.21 &-1.35 \\
$G_{\rm N1}$         &-1.47 & 0.91 &-0.68 &      &-1.79 &-0.25 \\
$G_{\rm N4}$         &-0.11 &-0.10 &-0.14 &-0.11 &      &      \\
$G_{\rm Y2}$         &      &-3.35 &      &      &-4.71 &-3.32 \\
\hline
\end{tabular}
\end{center}
\end{table}
%
% Table 2
%
\begin{table}[hbt]
\caption{Coupling constants used in the isobaric models for the
photo- and electro-production of kaons. Some of the sets of the
coupling constants were multiplied by -1. to get the negative sign
at $G_{\rm K\Lambda p}$ . } \label{models2}
\smallskip
\begin{center}
\begin{tabular}{|lrrrrrr|}
\hline
%\vspace{1mm}
\multicolumn{1}{|c}{Model}         &\multicolumn{1}{c}{AW3 \cite{AW88}}&
\multicolumn{1}{c}{AW4 \cite{AW88}}&\multicolumn{1}{c}{C3 \cite{WJC91}} &
\multicolumn{1}{c}{C4 \cite{WJC92}} &\multicolumn{1}{c}{SL \cite{SL96}} &
\multicolumn{1}{c|}{SLA \cite{SLA98}} \\
\hline
$G_{\rm K\Lambda p}$&-2.90 &-3.15 &-1.16 &-2.38 &-3.16 &-3.16 \\
$G_{\rm K\Sigma p}$ &-3.40 &-1.66 & 0.09 & 0.27 & 0.91 & 0.78 \\
$G^{V}_{\rm K^*}$   & 0.02 &-0.03 &      &-0.16 &-0.05 &-0.04 \\
$G^{T}_{\rm K^*}$   & 0.19 & 0.19 &      & 0.09 & 0.16 & 0.18 \\
$G^{V}_{\rm K1}$    &-0.10 &-0.13 &      & 0.02 &-0.19 &-0.23 \\
$G^{T}_{\rm K1}$    &-0.12 &-0.06 &      & 0.17 &-0.35 &-0.38 \\
$G_{\rm N1}$        & 4.49 & 1.11 &      &      &-0.02 &      \\
$G_{\rm N4}$        &-0.54 &-0.10 &-0.13 &-0.06 &      &      \\
$G_{\rm N6}$        &      &      &-0.25 &-0.09 &      &      \\
$G^1_{\rm N7}$      &      &      &      &      &-0.04 &-0.04 \\
$G^2_{\rm N7}$      &      &      &      &      &-0.14 &-0.12 \\
$G^1_{\rm N8}$      &      &      &      &      &-0.63 &      \\
$G^2_{\rm N8}$      &      &      &      &      &-0.05 &      \\
$G_{\rm Y1}$        &      &      & 0.13 &      &-0.42 &-0.40 \\
$G_{\rm Y2}$        & 0.29 &-0.70 &      &      & 1.75 & 1.70 \\
$G_{\rm L5}$        &      &      &      &      &-1.96 &-6.06 \\
$G_{\rm S1}$        &      &      &      &      &-7.33 &-3.59 \\
\hline
\end{tabular}
\end{center}
\end{table}

In Table 3 notation and parameters of the resonances are 
given because different notations are used in the literature 
which might cause a misunderstanding when a comparison of 
the model parameters is made with those in the original paper. 
%
% Table 3
%
\begin{table}[hbt]
\caption{Notation and parameters of the resonances 
considered by the models.} \label{reson}
\smallskip
\begin{center}
\begin{tabular}{|rlrl|}
\hline
%\vspace{1mm}
\multicolumn{2}{|c}{Nucleon resonances} & 
\multicolumn{2}{c|}{Hyperon resonances} \\
\hline
N1:& N$^*(1440)$ $P_{11} (\frac{1}{2})(\frac{1}{2}^+)$&
Y1(L1):& $\Lambda^*(1405)$ $S_{01} (0)(\frac{1}{2}^-)$\\
N4:& N$^*(1650)$ $S_{11} (\frac{1}{2})(\frac{1}{2}^-)$&
Y2(L3):& $\Lambda^*(1670)$ $S_{01} (0)(\frac{1}{2}^-)$\\
N6:& N$^*(1710)$ $P_{11} (\frac{1}{2})(\frac{1}{2}^+)$&
L5:& $\Lambda^*(1810)$ $P_{01} (0)(\frac{1}{2}^+)$\\
N7:& N$^*(1720)$ $P_{13} (\frac{1}{2})(\frac{3}{2}^+)$&
S1:& $\Sigma^*(1660)$ $P_{11} (1)(\frac{1}{2}^+)$\\
N8:& N$^*(1675)$ $D_{15} (\frac{1}{2})(\frac{5}{2}^-)$&
L8:& $\Lambda^*(1890)$ $P_{03} (0)(\frac{3}{2}^+)$\\
\hline
\end{tabular}
\end{center}
\end{table}

Tables 1 and 2 show that a common feature of the models, 
apart from C3, is that besides the extended Born diagrams 
they include also the vector kaon resonant one, K$^*$(890). 
Moreover, most of the models include also the  
axial-vector resonance K$_1$(1270). 
It was shown in Ref. \cite{WJC92} 
that these t-channel resonant terms in combination with 
s- and u-channel exchanges improve an agreement with data 
in the intermediate energy region. 
The models differ in a particular choice of the nucleon and 
hyperon resonances. Whereas most of the models assume only 
the baryon resonances with spin 1/2, in the Saclay-Lyon 
model (SL) the s-channel spin 3/2 (N7) and 5/2 (N8) nucleon 
resonances were included in addition, inconsistently with 
duality, to improve an agreement with data at higher 
energies \cite{SL96}.
The higher-spin (spin$>1/2$) s-channel resonances
were excluded in the C4 model motivated by the duality 
hypothesis.
The model SLA \cite{SLA98} is a simplified version of 
the full Saclay-Lyon model in which the nucleon resonance 
with spin 5/2 was left out. Predictions of the both models, 
however, are very similar for the cross sections and 
polarizations in the kaon photo-production \cite{SLA98}. 
Due to a complexity we do not include in the comparison 
more elaborate versions of the Saclay-Lyon model, B, C, 
and D \cite{SLA98}, which are based on the so-called 
off-shell extension inherent to the baryon resonances with 
spin$\geq$3/2. The N7 resonance was assumed also in 
the AB3 model \cite{ABW85} but we omit this model here  
too.

Free parameters of the models were obtained by a 
least-squares fit to various sets of experimental data 
which indicate an expected range of validity of the 
models. The models listed in Tab.1 and AW4 were 
confined to the kaon photo-production data for
$E_{\gamma}^{lab}<1.5$~GeV whereas  C4, SL, and SLA 
models were fitted to the data in the energy range up 
to $\approx 2$~GeV. The models in Tab. 2, except for AW3, 
were fitted also to electro-production data  
and some of them (C4, SL, and SLA) to the data on 
the radiative capture of K$^-$ on the proton in addition. 
The model AW3 was fitted only to electro-production 
data \cite{AW88}.

Some of the models violate the crossing principle, e.g. 
the AS1 over-predicts the branching ratio of the 
radiative capture \cite{AS90} but the models C4, SL and 
SLA keep the proper prediction for the ratio. 
The models AS1, AS2, AW2, AW4, SL, and SLA fulfil the SU(3) 
symmetry limits for the two main coupling constants 
or at least for one of them, as can be seen from Tables~1  
and 2, whereas the models AB2, W1, W2, AW3, C3, and C4 
violates them. The C4 model violates the SU(3) symmetry 
constraint for both the main coupling constants: 
$g_{\rm K\Lambda p}/\sqrt{4\pi} = -2.38$ and
$g_{\rm K\Sigma p}/\sqrt{4\pi} = 0.27$.

The Regge model was intended to describe the 
photo-production data for $E_{\gamma}^{lab}> 4$~GeV. 
Its extension to the electro-production reaction and  
to the energies as low as $W=1.8$~GeV was performed by 
multiplying contributions from the Regge trajectories 
with simple dipole form factors. The cut-off masses 
were chosen to be equal for both trajectories, 
$\Lambda_{\rm K}$ = $\Lambda_{\rm K^*}$, and set to 
fit the new experimental data \cite{Nic98}. 
The induced value of the cut-off mass, 
$\Lambda_{\rm K}^2$=1.5~GeV$^2$, is quite large 
resulting in a rather flat form factor and therefore 
in a very small corresponding effective kaon charge 
radius \cite{Rgg97}. Predictions of the model for 
the separated cross sections are, however, sensitive 
both to a value of the cut-off masses and to a shape 
of the form factor. 

The isobaric model by Bennhold {\it et al.} \cite{Ben99} 
(MAID in the following) includes besides the Born and K-meson 
resonant terms (K$^*$,K$_1$) the four s-channel resonances, 
$S_{11}(1650)$, $P_{11}(1710)$, $P_{13}(1720)$, and 
$D_{13}(1895)$ which have a significantly big branching ratio 
to the K$\Lambda$ channel. The resonance $D_{13}(1895)$, not 
yet discovered experimentally but predicted by 
the constituent quark model \cite{QM} (with the mass of 1960 
MeV), was added to explain the bump structure of the SAPHIR 
data on the total cross section \cite{SPH98} around 
$E^{lab}_{\gamma}=1.5$~GeV ($W=1.9$~GeV). 
However, in Ref. \cite{Hampt} it was shown that the bump 
structure can be equally well reproduced when two hyperon  
resonances, L5 and L8 (see Tab. 3), are introduced instead 
of the ``missing'' $D_{13}(1895)$ resonance and the off-shell 
effects are introduced for the spin 3/2 resonances \cite{SLA98}.
The gauge-invariant prescription by Haberzettl \cite{HBMF98} 
was utilised to introduce the hadron form factors in the MAID 
model. Phenomenological parametrisation for the hadron form 
factors was assumed and different values of the cut-off mass 
were allowed for the background (Born, K$^*$, and K$_1$) and 
resonant contributions. The cut-off masses together with 
the couplings and the mass of the $D_{13}$ resonance were 
fixed by the photo-production data. Parameters of the 
electro-magnetic form factors of the resonances were 
fitted \cite{Ben99} to the electro-production 
data \cite{Nic98}. 
We used the interactive code at the KAON-MAID 
Web-page \cite{Ben99} to generate numerical results of 
the model. We made no scaling or any other changes of the 
original model. Our results are slightly different 
from those in Refs. \cite{HBMF98} and \cite{Ben99} but 
they differ more distinctly from the results presented in 
Ref. \cite{Moh02}(Fig.~16 for $\sigma_L$) where, however, 
other code was used than that from the Web-page (see the 
reference 8 in \cite{Moh02}).     

\section{Discussion of predictions of the models}

In this section we discuss predictions of the models in view 
of the latest experimental data on the \PP and \EP reactions. 
A similar analysis of the models can be also found in 
Ref. \cite{Tohoku}, applied particularly to the photo-production 
reaction and focused at energies around 
$E_{\gamma}^{lab}=1.3$~GeV and small kaon angles because it 
aimed at a selection of models appropriate for calculations 
of the hyper-nuclear production cross sections. 

\subsection{Electro-production}

\begin{figure}[htb]
\centerline{\psfig{figure=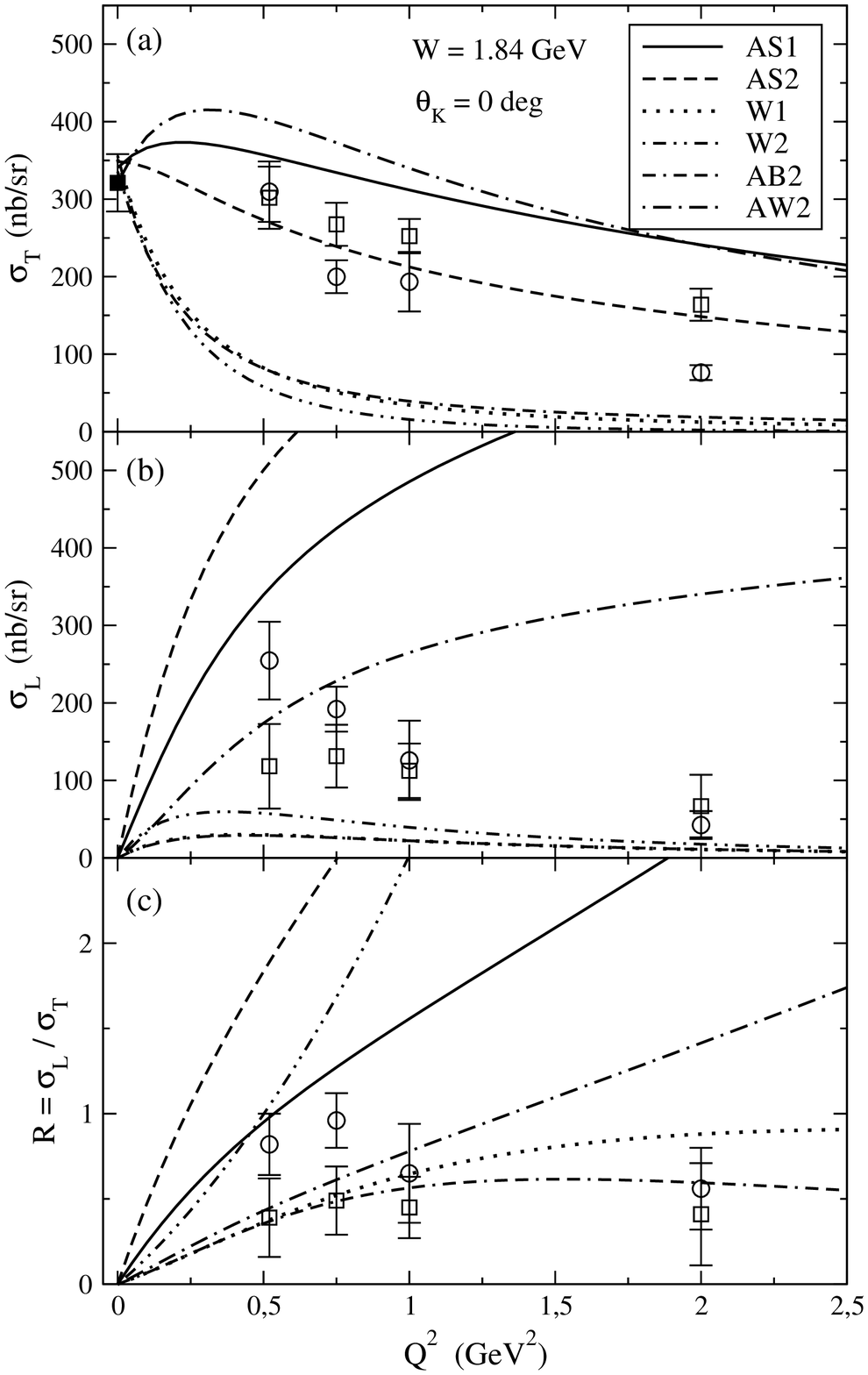,width=12.cm}}
%\vspace*{-5mm}
\caption{Separated cross sections $\sigma_T$ and $\sigma_L$ 
and their ratio $R$ are shown as a function of $Q^2$ at 
$W=1.84$~GeV and zero kaon c.m. angle. Predictions of the 
models listed in Table 1 are plotted. The data are from 
Refs. \cite{Nic98} (circle), \cite{Moh02} (square), 
and \cite{Ble70} (solid square).}
\label{r1a}
\end{figure}

\begin{figure}[htb]
\centerline{\psfig{figure=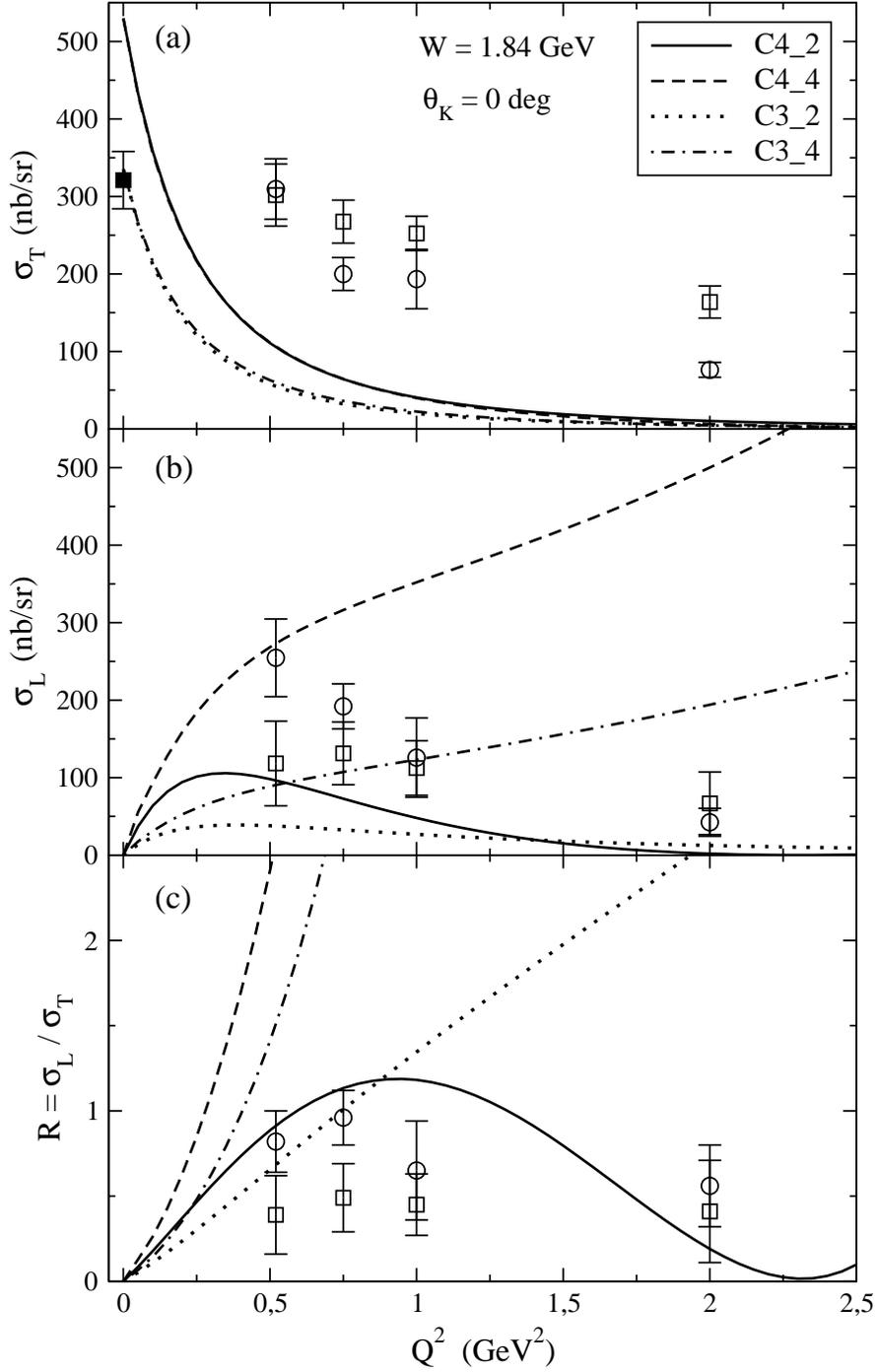,width=12.cm}}
%\vspace*{-5mm}
\caption{The same as in Fig.1 but for Williams-Ji-Cotanch 
models with different prescriptions for the electro-magnetic 
form factors (see sect. 2).}
\label{r1b}
\end{figure}

\begin{figure}[htb]
\centerline{\psfig{figure=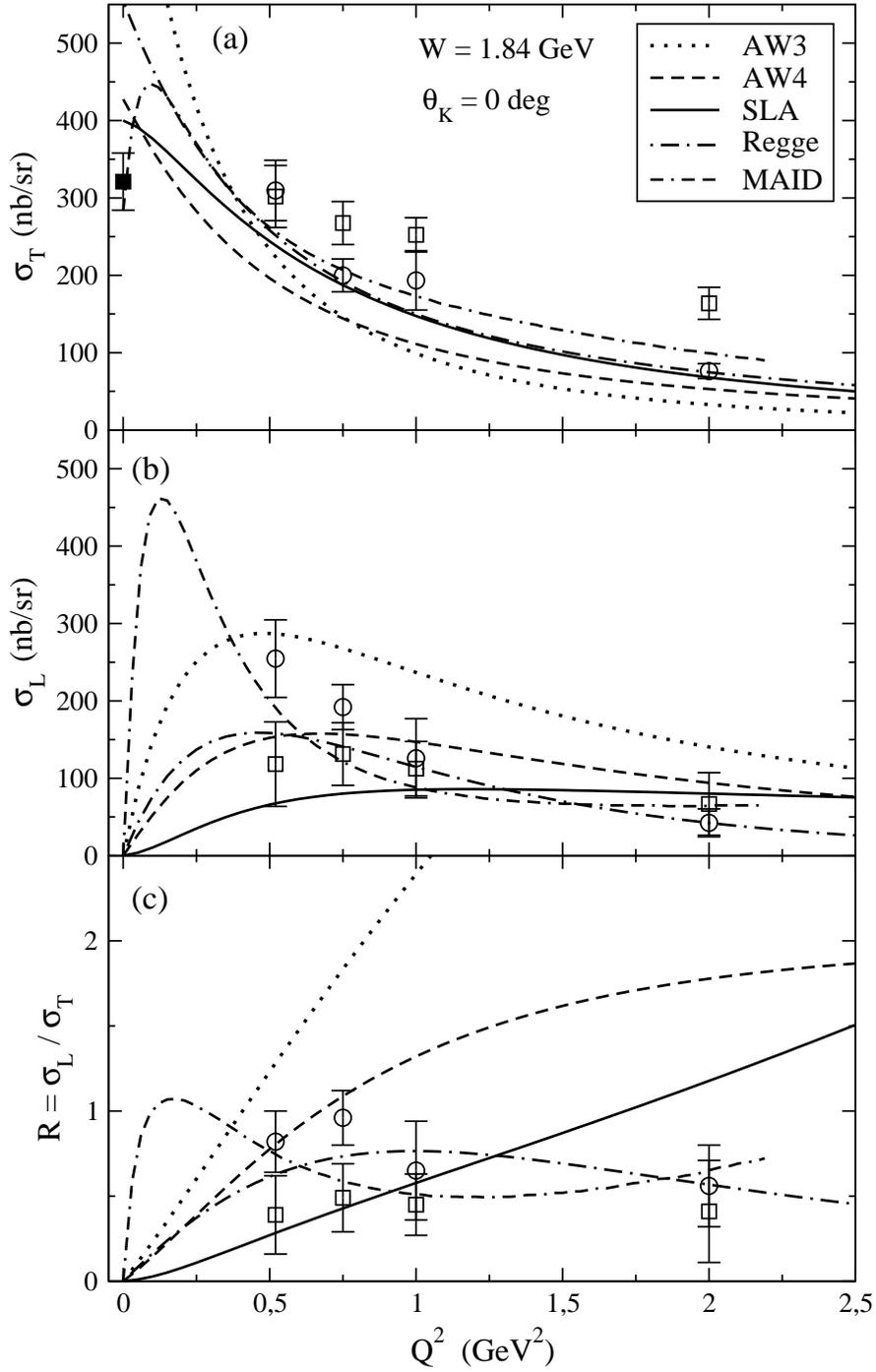,width=12.cm}}
%\vspace*{-5mm}
\caption{The same as in Fig.1 but for AW3, AW4, SLA, Regge, 
and MAID models.}
\label{r1c}
\end{figure}

In Figures~\ref{r1a}-\ref{r1c} we compare predictions 
of the isobaric and Regge models for \EP with two sets 
of new experimental data. The two data sets, 
Niculescu {\it et al.} \cite{Nic98} and Mohring 
{\it et al.} \cite{Moh02}, are results of two different 
analysis of the experiment E93-018 performed at JLab. 
Details of the differences of the two analysis are reported 
in subsection D of Ref. \cite{Moh02}. For the purposes of 
the present paper, in comparing data with the models, we 
have chosen to report on both results of analysis.  
The separated longitudinal 
($\sigma_L$) and transverse ($\sigma_T$) cross sections 
and their ratio $R=\sigma_L / \sigma_T$ are plotted
for zero kaon c.m. angle ($|t|=|t|_{min}$) as a function 
of $Q^2$. The photo-production point at 
$E_{\gamma}^{lab}=1.3$~GeV ($W=1.82$~GeV) and 
$\theta_K^{cm}=6$~deg \cite{Ble70} is also shown for 
$\sigma_T$.  

Most of the extended models, AS1, W1, W2, AB2, and AW2 
provide poor results for both $\sigma_T$ and $\sigma_L$.  
Only the AS2 model is in a very good agreement with the 
latest re-analysed data \cite{Moh02} for $\sigma_T$ 
but it fails for $\sigma_L$  and $R$ (Fig.~\ref{r1a}).  
The models AB2 and W1 give almost identical results 
for $\sigma_T$ and $\sigma_L$ which can be understood 
from Table~1. The two models include the same resonances 
and the values of the corresponding coupling constants are 
close each other except for $G_{\Sigma}$ whose contribution 
is small here. 
In the AS1, AS2, and AW2 models the K$_1$ and Y$_2$ resonances 
are added which results in a qualitatively different 
$Q^2$-dependence of both $\sigma_L$ and $\sigma_T$ which 
is more pronounced at $Q^2<0.5$~GeV$^2$. This difference is 
caused mainly by a contribution of the K$_1$ exchange 
in the AS1, AS2, and AW2 models. 
The bottom part of Fig.~\ref{r1a} demonstrates that 
even though the models fail in description of the separated 
cross sections some of them can provide a very good 
description of the ratio $R$ up to $Q^2=1$~GeV$^2$ 
(AW2 and W1) and the AB2 model even up to 2~GeV$^2$. 
All the six models (Table~1) give very good 
predictions for the photo-production cross section  
(Fig.~\ref{r1a}, the top part).

In Figure~\ref{r1b} we demonstrate sensitivity of 
predictions of the C3 and C4 models to procedure  
of inclusion of the electro-magnetic form factors. 
Whereas, there is almost no sensitivity for $\sigma_T$,
the longitudinal cross section and the ratio $R$ exhibit 
a very big differences, especially for the C4 model. 
Utilising the original prescription in which the 
electric ($G_E$) and magnetic ($G_M$) form factors are 
substituted in the electro-magnetic vertexes (C3\_4 and 
C4\_4), leads to a wrong $Q^2$-dependence of $\sigma_L$ 
and $R$ for both models. In our further calculations we 
will utilise the second choice ($F_1$ and $F_2$ form 
factors are used, C3\_2 and C4\_2) because it also fits 
the original calculations \cite{WJC92} much better than 
the original prescription does. This choice is equivalent 
to that used also in Ref. \cite{SL96}. 
However, even this choice cannot describe the data
satisfactorily. Some, more elaborate prescription for 
the form factors is, probably, needed. For results of 
the ``latest'' version of the C4 (WJC) model, which 
differ from our results, we refer to Fig.~16 in 
Mohring {\it et al.} \cite{Moh02}.
The simpler C3 model which do not include neither 
K$^*$ nor K$_1$ resonance (Table~2) provides  
bad results for both $\sigma_T$ and $\sigma_L$ 
(the dotted line in Fig.~\ref{r1b}).

In Figure~\ref{r1c} results of more elaborate models, 
Regge, MAID and those presented in Table~2 are 
revealed. They were fitted to both photo- and 
electro-production data, except for the model AW3 
which was fitted to \EP data only. The results of 
these models for $\sigma_T$ are consistent with 
the analysis by Niculescu {\it et al.} \cite{Nic98} 
rather than with the re-analysis by Mohring 
{\it et al.} \cite{Moh02}. The results for $\sigma_L$, 
excluding those of AW3, on the contrary, are in 
a better agreement with the analysis by 
Mohring {\it et al.}. Predictions of the models 
differ significantly each 
other for both cross sections at $Q^2<0.5$~GeV$^2$, 
especially those of the MAID model for $\sigma_L$ 
which follow, as expected, the older data \cite{Nic98} 
at small $Q^2$ shaping a pronounced bump structure at 
$Q^2\approx0.2$~GeV$^2$. 
As discussed in Ref. \cite{Ben99} the bump structure 
of the longitudinal and transverse cross sections 
is created by the Dirac electro-magnetic form factor 
($F_1$) of the ``missing'' $D_{13}(1895)$ resonance 
which was determined by the electro-production data 
\cite{Nic98} to be very steep at small $Q^2$.  
More precise data for $Q^2<0.5$~GeV$^2$ 
would be therefore useful to supply additional 
constraints on the models. Good results are also 
provided by the AW4 model for $\sigma_L$ and $R$ for 
$Q^2<0.75$~GeV$^2$.  
The ratio $R$ is described satisfactorily only by the 
Regge and MAID models. The SLA model provides a good 
description for the first three data points of 
Mohring {\it et al.} but fails for larger $Q^2$. 
The AW4 and AW3 models over-estimate the data of 
Ref. \cite{Moh02} and give a wrong $Q^2$-dependence 
of $R$ for $Q^2>1$~GeV$^2$.
The model AW3 is known to provide wrong results for 
the \PP reaction which can be seen in the top part 
of Fig.~\ref{r1c} where AW3 exceeds the 
photo-production cross section by factor of 3. 
The Regge (Fig.~\ref{r1c}) and C4 (Fig.~\ref{r1b}) 
models also over-predict $\sigma_T$ at $Q^2\approx 0$,  
C4 giving by $\approx 100$~nb/sr larger values than 
the SLA and AW4 models. 
This observation is in agreement with conclusions 
drawn in Ref. \cite{Tohoku} where it was shown that 
C4 (WJC) considerably over-predicts the experimental 
data at zero kaon angle and at $E_{\gamma}^{lab}=1.3$~GeV. 
The only MAID model provides an excellent agreement 
with the photo-production point. However, it should be 
further investigated whether the bump structure 
exhibited by the MAID model at $Q^2\approx 0.2$~GeV$^2$ 
is inherent to the dynamics of the process, for example 
a presence of the $D_{13}(1895)$ resonance,  or if 
the inconsistency of the photo- and electro-production 
data is of some other origin, e.g. a problem of 
data normalisation. Precise data on the separated cross 
sections at $0<Q^2<0.5$~GeV$^2$ could help to clarify  
this point.    

In addition to that, predictions of the models for the 
unseparated cross section, 
$\sigma_{UL}=\sigma_T + \epsilon \;\sigma_L$, were 
compared with experimental results as they are given 
in Ref. \cite{Moh02}.
In Table 4 we list values of the $\chi^2$ function 
which was calculated for the twelve data 
points \cite{Moh02} as follows\\
\begin{equation}
 \chi^2 = \frac{1}{12} \sum_i \left(\frac{\sigma_{UL}^{th}
- \sigma_{UL}^{exp}}{\Delta\sigma_{UL}^{exp}}\right)^2\;\; .
\end{equation}
For a better comparison values of $\chi^2$ were 
normalised to unity for the AW3 model and listed 
as $\chi_N^2$. The best results on $\sigma_{UL}$ 
were provided by AW3, AW4, Regge, SLA, and C4\_4 
models which in the case of the AW3 and C4\_4 models 
obviously contradicts to our previous observations
made in Figs.~\ref{r1b} and \ref{r1c}. 
These two models under-predict  $\sigma_T$ and 
simultaneously over-predict $\sigma_L$ data which 
results in a better agreement with the $\sigma_{UL}$ 
cross sections.
Moreover, both models fail in predicting $R$ for 
large $Q^2$. Only the Regge, SLA, and AW4 models 
can predict acceptable values for both the separated 
and unseparated cross sections simultaneously. 
The C4\_2, C3 and the extended models (Tab.1)
provide much worse results for the $\chi^2$, see 
Table 4.
%
% Table 4
%
\begin{table}[hbt]
\caption{The $\chi^2$ calculated for the unseparated 
cross section, using Eq.(1) with experimental points 
taken from Ref. \cite{Moh02}. The values $\chi^2_N$ 
are normalised to the value of $\chi^2$ for the AW3 
model. \label{chi}}
\smallskip
\begin{center}
\begin{tabular}{|lrr||lrr|}
\hline\vspace{1mm} Model & $\chi^2$ & $\chi_N^2$ & 
Model & $\chi^2$ & $\chi_N^2$\\
\hline\vspace{1mm}
AB2   &  599.4 &  9.35  &  AW3   &  64.1 &  1.00\\
AW2   &  391.2 &  6.10  &  AW4   & 116.9 &  1.82\\
W1    &  617.5 &  9.63  &  C3\_2 & 678.9 & 10.59\\
W2    &  652.4 & 10.18  &  C3\_4 & 386.6 &  6.03\\
AS1   & 1069.6 & 16.69  &  C4\_2 & 502.5 &  7.84\\
AS2   & 1141.8 & 17.81  &  C4\_4 &  70.5 &  1.10\\
Regge &   89.7 &  1.40  &  SLA   & 117.2 &  1.83\\
\hline
\end{tabular}
\end{center}
\end{table}

\begin{figure}[htb]
\centerline{\psfig{figure=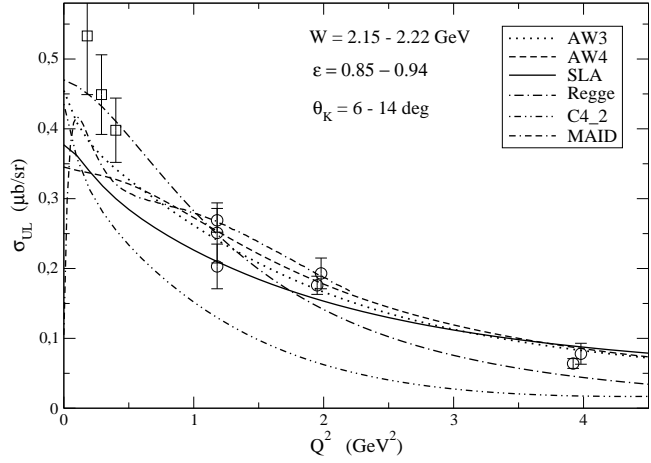,width=11.cm,angle=270}}
%\vspace*{-5mm}
\caption{Unseparated cross section for \EP is shown as 
a function of $Q^2$. Calculations were performed at 
$W=2.18$~GeV, $\epsilon =0.91$, and $\theta_K^{cm}=11$~deg. 
The best predictions were shown only. The data are from 
Refs. \cite{Beb77} (circle) and \cite{Bro72} (square).}
\label{r6b}
\end{figure}

\begin{figure}[htb]
\centerline{\psfig{figure=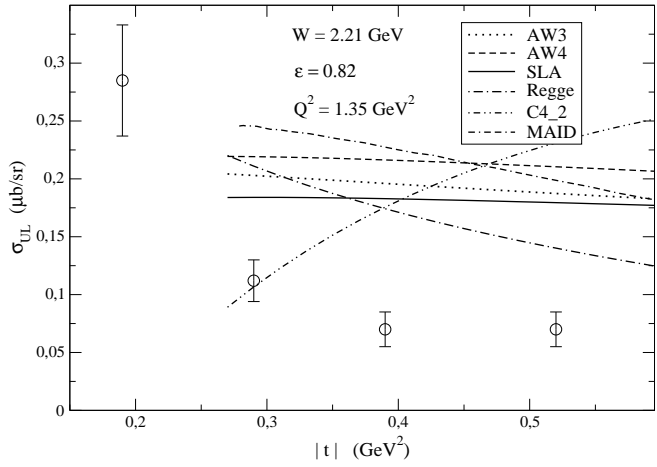,width=11.cm,angle=270}}
%\vspace*{-5mm}
\caption{Unseparated cross section is shown as a function of 
the momentum transfer $|t|$. The physical region starts 
 at $|t|_{min}=0.26$~GeV$^2$. The data are from 
Ref. \cite{Bra79}.}
\label{r6c}
\end{figure}

To reveal behaviour of the models in the kinematic 
region relevant for the E98-108 experiment, 
$W=1.8 - 2.2$~GeV and $Q^2=1.9 -2.4$~GeV$^2$, 
in Figs.~\ref{r6b} and \ref{r6c} we compare predictions 
of the models for the unpolarised cross section 
$\sigma_{UL}$ with old data by Bebek 
{\it et al.} \cite{Beb77}, Brown {\it et al.} \cite{Bro72}, 
and Brauel {\it et al.} \cite{Bra79}.
In Figure~\ref{r6c} the data point at 
$|t|=0.19$ GeV$^2$ \cite{Bra79} is out of the physical 
region, for $W=2.21$ GeV starting at 
$|t|_{min}=0.26$ GeV$^2$, which is probably due to 
scaling the data from larger 
energies ($|t|_{min}=0.19$~GeV$^2$ for $W=2.43$~GeV).  
In Figure~\ref{r6b} calculations were performed at 
$W=2.18$~GeV, $\epsilon=0.91$, and 
$\theta_{K}^{cm}=11$~deg. 
Only the best predictions are shown in the both 
figures. The models AW3, AW4, SLA, MAID, and Regge can 
provide very good results for the $Q^2$-dependence, 
Fig.~\ref{r6b}. Results of the models differ 
mainly at small $Q^2$ again where more precise data 
would be useful. Here the MAID model displays 
a pronounced bump structure similar to that for 
the separated cross sections (Fig.~\ref{r1c}) but 
here it predicts very small, in comparison with 
the other models, photo-production cross section: 
$\sigma =105$ nb/sr (Fig.~\ref{r6b}).
The models fail in describing the t-dependence 
of $\sigma_{UL}$ the best results being achieved for 
$|t|_{min}$ (zero kaon angle), Fig.~\ref{r6c}.  
The only Regge and MAID models provide a reasonable 
behaviour but they fail in normalisation of 
$\sigma_{UL}$. Predictions of the model C4 are out 
of the data in both Figures \ref{r6b} and \ref{r6c} 
(excluding at $|t|_{min}$ in Fig. \ref{r6c}). 
The model possesses a wrong t-dependence 
in addition and its results differ distinctly from 
predictions of the other models at forward kaon 
angles (Fig.~\ref{r6c}).

\subsection{Photo-production}
In this subsection we discuss only the isobaric 
models because the energies assumed here are not high 
enough to be reasonable to apply the Regge model. 
The Regge model systematically over-predicts the 
photo-production data at photon energies less than 
2~GeV as was shown in Fig.~\ref{r1c} and Ref. \cite{Tohoku}. 

\begin{figure}[htb]
\centerline{\psfig{figure=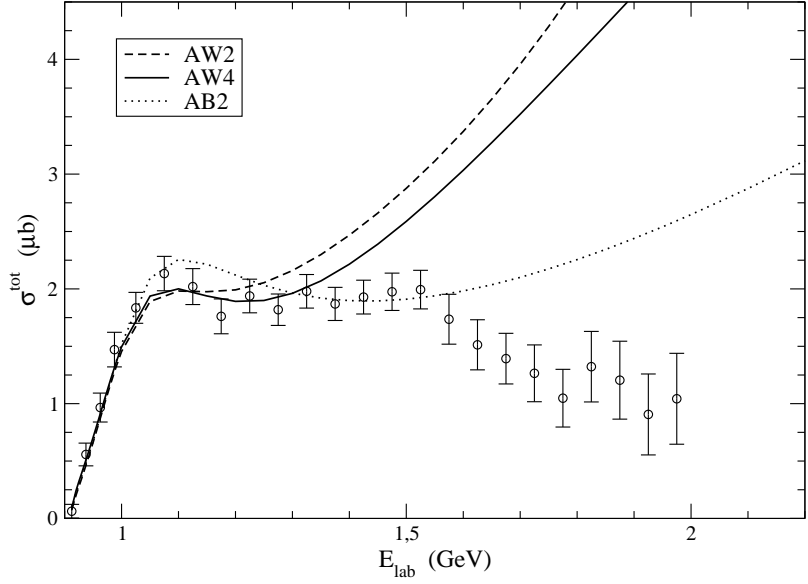,width=11.cm,angle=270}}
%\vspace*{-5mm}
\caption{Total cross sections for \PP as predicted by AW2, 
AW4, and AB2 models are shown in dependence of the photon 
energy. The data are from Ref. \cite{SPH98}.}
\label{r2a}
\end{figure}

\begin{figure}[htb]
\centerline{\psfig{figure=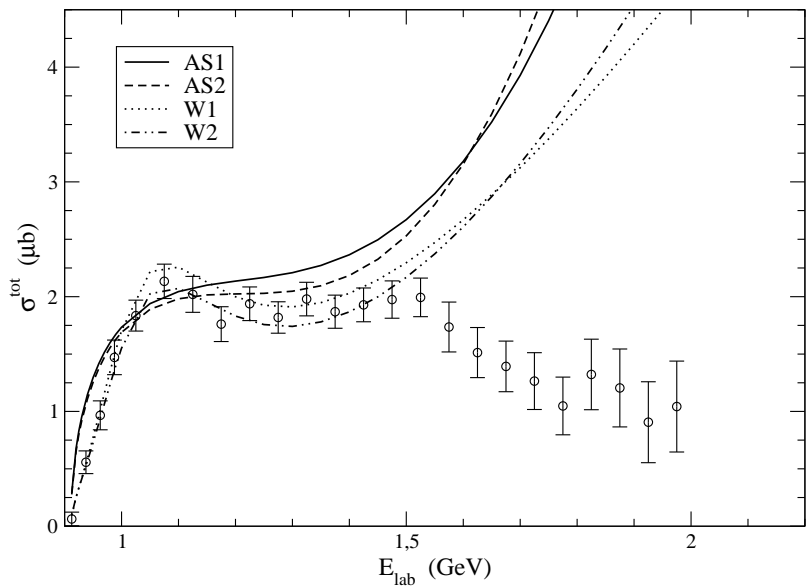,width=11.cm,angle=270}}
%\vspace*{-5mm}
\caption{The same as in Fig.~6 but for AS1, AS2, W1, and W2 
models.}
\label{r2b}
\end{figure}

\begin{figure}[htb]
\centerline{\psfig{figure=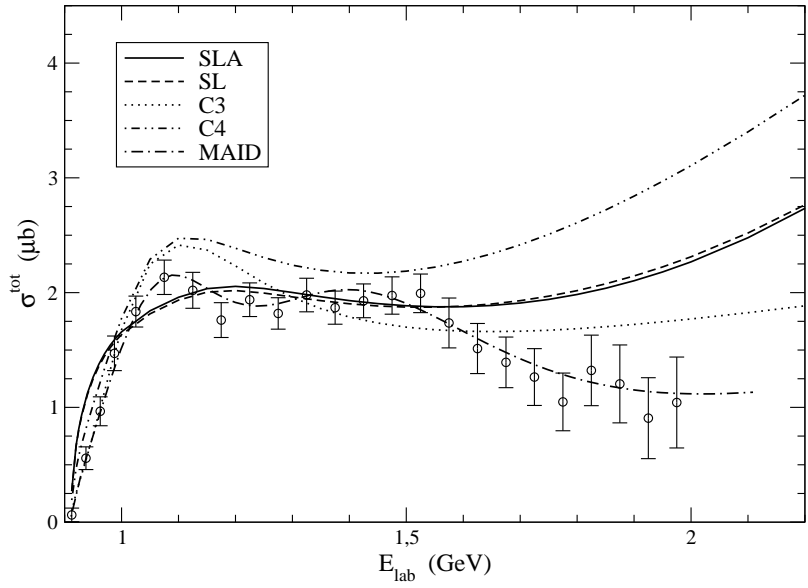,width=11.cm,angle=270}}
%\vspace*{-5mm}
\caption{The same as in Fig.6 but for SLA, SL, C3, C4, and 
MAID models.}
\label{r2c}
\end{figure}

In Figures \ref{r2a}-\ref{r2c} we compare predictions 
of the models for the total cross section in the 
photo-production with SAPHIR data \cite{SPH98}. 
Relative errors of the data amount approximately 
to 8\% excluding in the threshold region where it is 
around 15\% and for energies higher than 1.5~GeV 
where it amounts up to 40\% 
(see Figures \ref{r2a}-\ref{r2c}). These data are 
therefore less restrictive than the unseparated 
electro-production cross sections for which the 
relative errors are less than 5\%.

Predictions of the older models AS1, AS2, AW2, and AW4 
are consistent with the data only up to 
$E_{\gamma}^{lab}= 1.4$~GeV (Figs. \ref{r2a} and \ref{r2b})  
and those of W1, W2, AB2, and C4 up to 
$E_{\gamma}^{lab}= 1.5$ GeV (Figs. \ref{r2b} and \ref{r2c}). 
Apart from the C4 model the agreement with the data 
is within the expected range of validity of the models.
The SL, SLA, and C3  models provide a satisfactory 
description of $\sigma^{tot}$ for the photon energy 
ranging up to 2 GeV (Fig. \ref{r2c}).
Predictions of the SL and SLA models are very 
close each other. Better results are provided by 
the ``off-shell extended'' versions B and C of the 
Saclay-Lyon model (see Ref. \cite{SLA98} (Fig.~2)). 
The MAID model is in an excellent agreement with the 
data in the whole range of energies describing also 
their dip-bump structure (Fig. \ref{r2c}). This success  
is due to presence of the hadron form factors in the 
model as well as to the fit of the model parameters to 
the described data. The model attempted to explain the bump 
structure of the data around $E^{lab}_{\gamma}=1.5$~GeV 
by addition the ``missing'' resonance $D_{13}(1895)$ 
to the model. Inclusion of this resonance also 
improved behaviour of model predictions at higher 
energies \cite{Ben99}.

Values of the $\chi^2$, calculated in analogy with Eq.(1) 
but for the total cross section and normalised to the 
number of the data points, are 2.47 and 5.17 for C3 and 
SLA models, respectively. Larger value for the SLA model 
is due to worse predictions of the model for $\sigma^{tot}$ 
near the threshold and for $E_{\gamma}^{lab}>1.6$ GeV 
than those of the C3 model.
The model AB2 gives a very good agreement with data 
up to 1.6 GeV (Fig. \ref{r2a}), resulting in 
$\chi^2=5.10$ which is better than that for SLA. 
However, prediction of the AB2 is worse for 
$E_{\gamma}^{lab}>1.6$~GeV (Figs. \ref{r2a} 
and \ref{r2c}) than that of SLA. 
The best value, $\chi^2=0.32$, was achieved by the MAID 
model whereas for C4  it amounts 11.6.  

The SAPHIR data on the differential cross sections and 
polarizations \cite{SPH98} suffer from still larger error 
bars than those for $\sigma^{tot}$. 
Another disadvantage of the data is their averaging over 
bins of the photon energy and kaon angle. 
At this point one should also mention an apparent discrepancy 
of these SPHIR data with the older ones \cite{SPH94}, 
especially at small kaon scattering angles \cite{Tohoku}.
However, we show a comparison of model predictions 
with the latest data here. 

\begin{figure}[htb]
\centerline{\psfig{figure=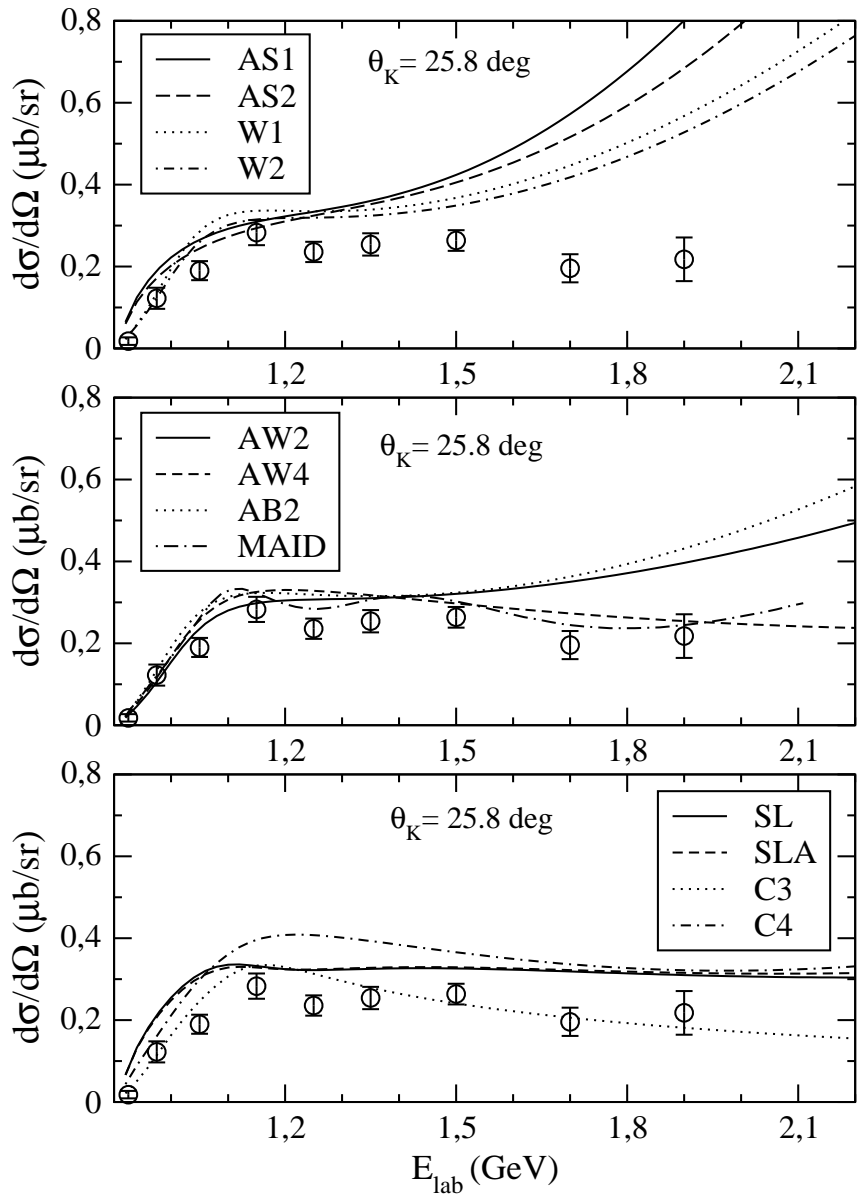,width=12.cm}}
%\vspace*{-5mm}
\caption{Differential cross section is shown as a function 
of the photon energy at kaon c.m. angle of 25.8 deg. 
The data are from Ref. \cite{SPH98}.}
\label{r3a}
\end{figure}

\begin{figure}[htb]
\centerline{\psfig{figure=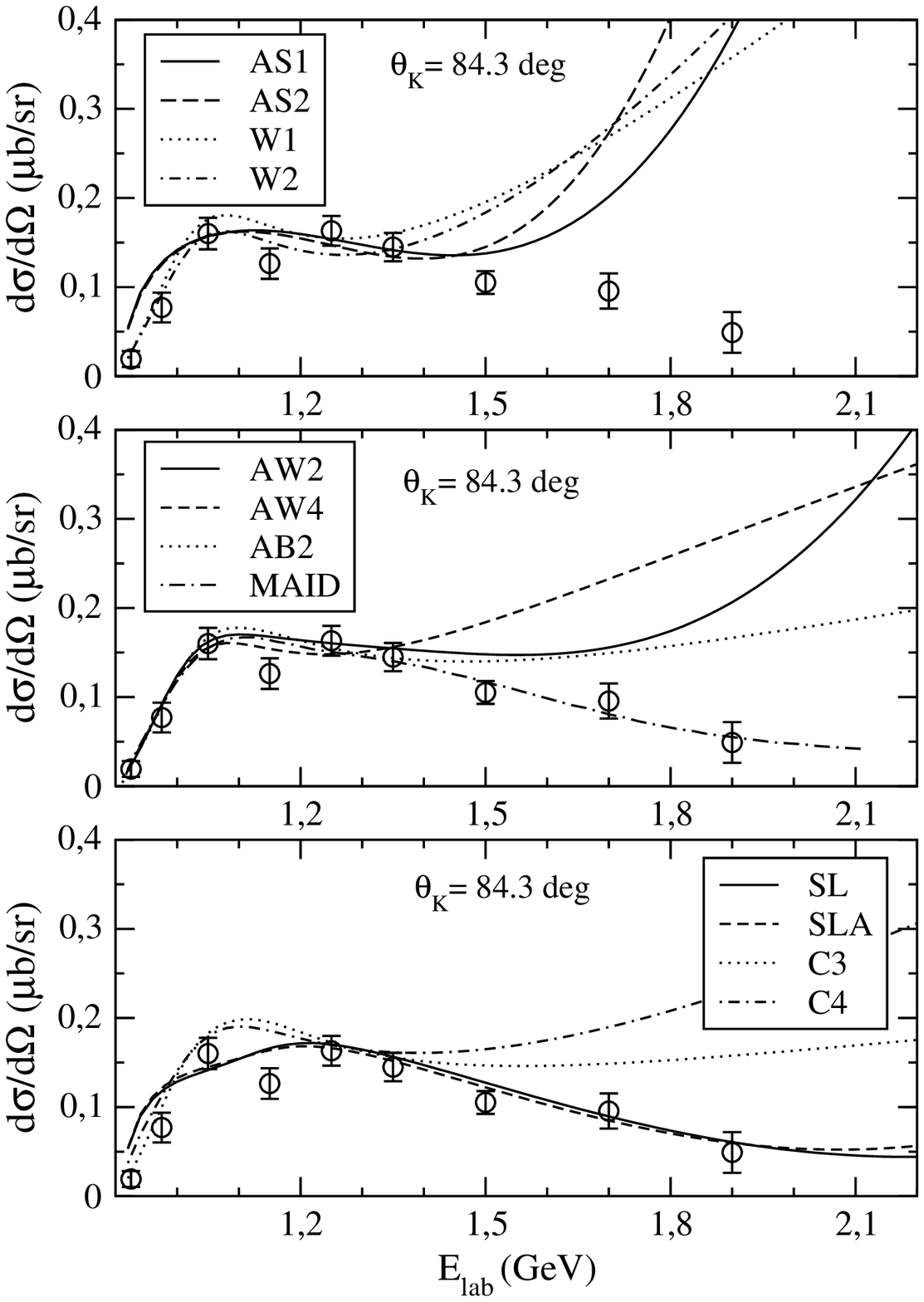,width=12.cm}}
%\vspace*{-5mm}
\caption{The same as in Fig.~9 but for kaon angle of 84.3 deg.}
\label{r3b}
\end{figure}

\begin{figure}[htb]
\centerline{\psfig{figure=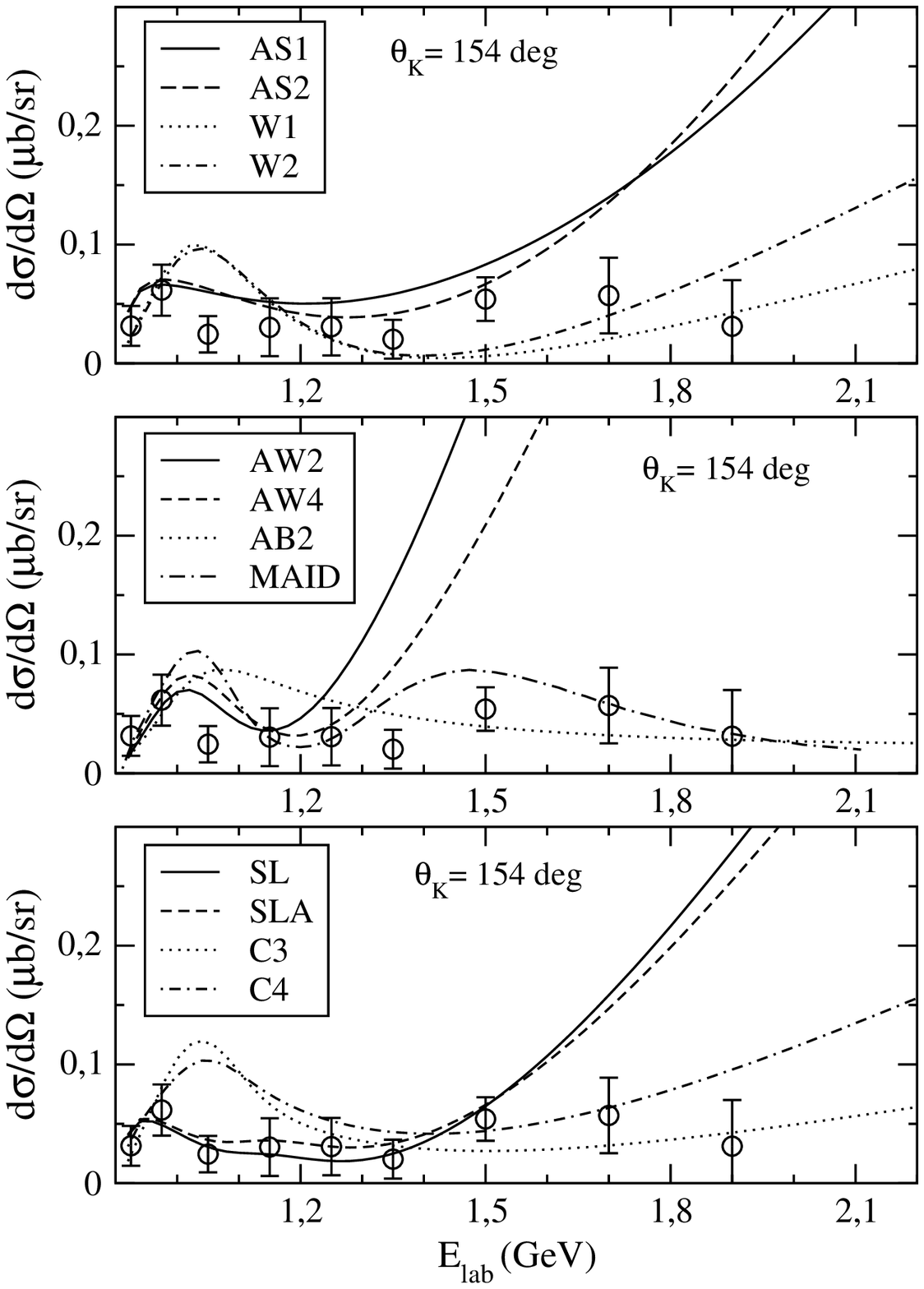,width=12.cm}}
%\vspace*{-5mm}
\caption{The same as in Fig.~9 but for kaon angle of 154 deg.}
\label{r3c}
\end{figure}

In Figs. \ref{r3a}-\ref{r3c} we display the differential 
cross section as a function of the photon lab energy 
for three values of the kaon c.m. angle. 
The older models AS1, AS2, W1, W2, AW2, AW4, and AB2 can 
describe the data satisfactorily only up to 1.4~GeV, 
even if the AB2, W1 and W2 models at 154 deg 
(Fig. \ref{r3c}) and the AW4 model at 25.8 deg 
(Fig. \ref{r3a}) can provide good results for 
energies up to 2 GeV. 
The models C3 and AW4 give very good results at small 
angles (Fig. \ref{r3a}) and high energies but they 
fail for larger angles and $E_{\gamma}^{lab}>1.4$~GeV 
(AW4) or $1<E_{\gamma}^{lab}<1.2$~GeV (C3) 
(Figs. \ref{r3b} and \ref{r3c}).
The model C4 systematically over-predicts the 
differential cross sections in most of the kinematic 
region discussed here, especially at forward angles 
which was already observed in Fig. \ref{r1b}. 
The SL and SLA models over-predict data at small 
angles especially near the threshold region. They 
provide very good description at intermediate angles 
but fail again at backward angles and 
$E_{\gamma}^{lab} > 1.5$~GeV (Fig. \ref{r3c}) where 
results of the two models differ noticeably too. 
Evidently, the best result here is provided by 
the MAID model which can reproduce even the data 
structure. The model being regularised by the hadron 
form factors agree with the data for all angles 
even for $E_{\gamma}^{lab}=1.9$~GeV. 

\begin{figure}[htb]
\centerline{\psfig{figure=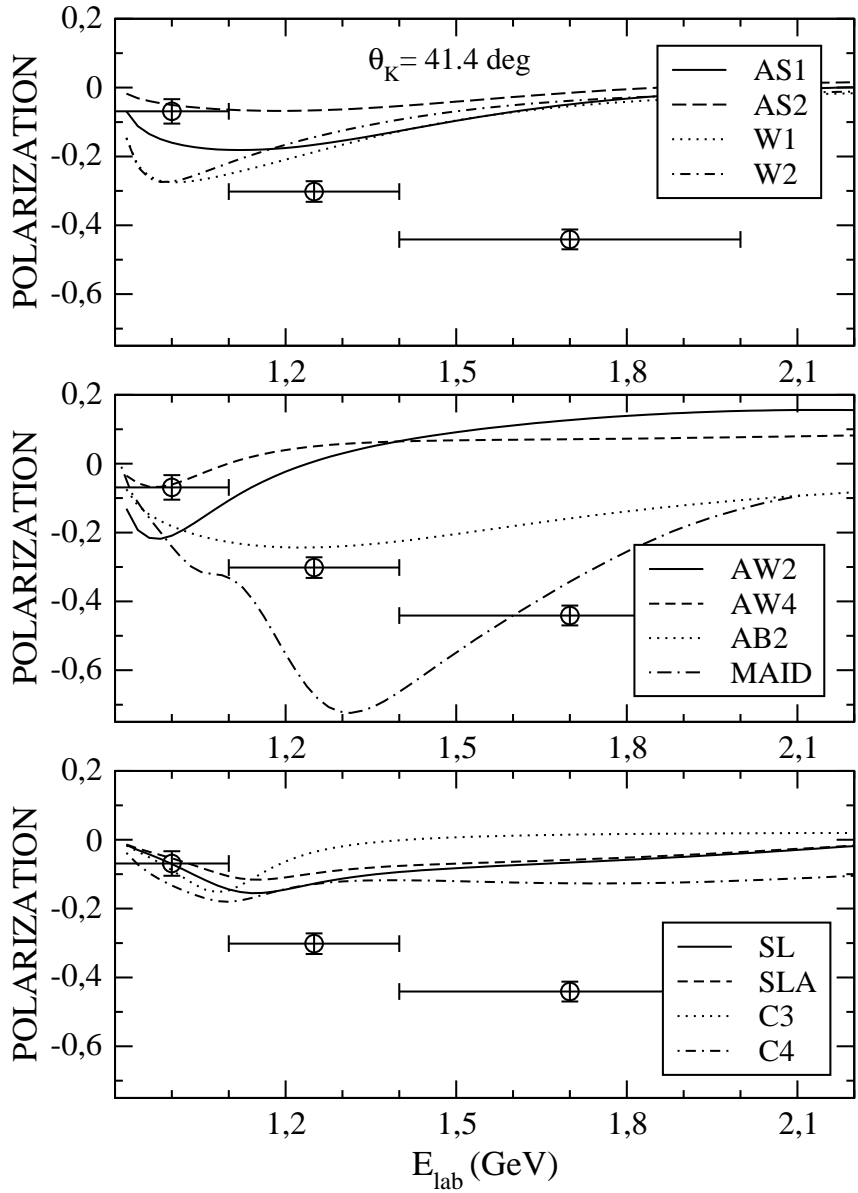,width=12.cm}}
%\vspace*{-5mm}
\caption{Lambda polarization asymmetry is shown as a function 
of the photon energy at kaon c.m. angle of 41.4 deg. 
The data are from Ref. \cite{SPH98}.}
\label{r3d}
\end{figure}

\begin{figure}[htb]
\centerline{\psfig{figure=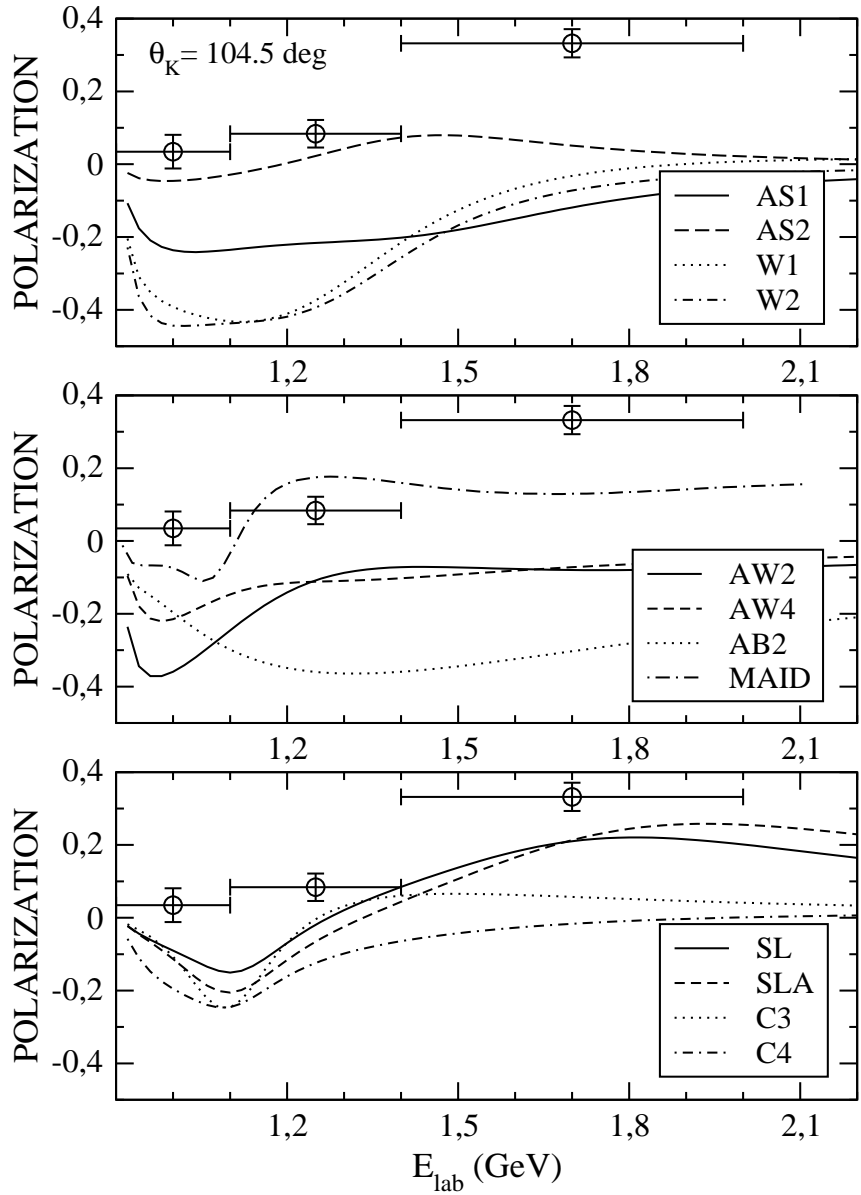,width=12.cm}}
%\vspace*{-5mm}
\caption{The same as in Fig.~12 but for kaon angle of 
104.5 deg.}
\label{r3e}
\end{figure}

Situation with data accuracy is still worse for 
the lambda polarization asymmetry ($P$). That is why 
in Figs. \ref{r3d} and \ref{r3e} we show the energy bins 
over which data were averaged. At $\theta_K^{cm}=41.4$ deg 
(Fig. \ref{r3d}) the models except for MAID have a problem to 
describe the data points for $E_{\gamma}^{lab} > 1.3$~GeV.  
The best result (in terms of $\chi^2$) was achieved 
by the MAID and AB2 models. At $\theta_K^{cm}=104.5$ deg 
(Fig. \ref{r3e}) the MAID, Saclay-Lyon, and AS2 (for 
$E_{\gamma}^{lab} < 1.4$ GeV only) models give reasonable 
predictions. 
There is no need to say that more and better quality 
data are desirable here.

\section{Conclusions}
Among the tested models only the Saclay-Lyon A (SLA) and 
MAID can describe more or less all the discussed 
data. 
The Regge model is successful in predicting the 
electro-production cross sections but it cannot provide 
good results for the photo-production at energy less than 
3-4 GeV. 
The Williams-Ji-Cotanch model C4 fails to describe 
the total cross section for the photo-production 
whereas C3 version cannot describe the new electro-production 
data. However, uncertainty in prescription for 
the electro-magnetic form factors does 
play a significant role for predictions of these models for 
the separated cross sections 
in the electro-production. 
The older models, Adelseck-Saghai (AS1, AS2), Workman (W1, W2), 
and Adelseck-Bennhold (AB2) are excluded by the new 
photo-production data, except for some cases where the models 
provide a reasonable results. These models, extended for 
the electro-production, cannot also predict the new data 
satisfactory. The analysis depends, however, to some extent 
on the prescription used for the electro-magnetic 
form factors. 
Large differences of the model predictions for the separated 
cross sections at small $Q^2$ call for other high accuracy 
electro-production data in that region.

In the kinematic region of the E98-108 experiment, 
$W=1.8 - 2.2$~GeV and $Q^2=1.9 - 2.4$~GeV$^2$, the SLA, 
Adelseck-Wright (AW4), Regge and MAID models are
expected to provide reasonable results for the separated 
and unseparated  cross sections whereas the unseparated 
cross sections merely can be predicted by the AW3 and C4 
models in addition, where in the model C4 the electric 
($G_E$) and magnetic ($G_M$) form factors are substituted 
in the electro-magnetic vertex \cite{WJC92}.

\section{Acknowledgements}
Financial support from GA CR 202/02/0930 is gratefully 
acknowledged. Two of us (P.B. and M.S.) thank the INFN, 
Gruppo Sanit\`{a} in Rome sincerely for its hospitality 
during completing this work.

%===========================================================================

%
% Figure 1
%
%
% Figure 2
%
%
% Figure 3
%
%
%
% Figure 4
%
%
%
% Figure 5
%
%
%
% Figure 6
%
%
%
% Figure 7
%
%
%
% Figure 8
%
%
%
% Figure 9
%
%
%
% Figure 10
%
%%
% Figure 11
%
%%
% Figure 12
%
%
%
% Figure 13
%

\end{document}